*Article*

# The Societal Implications of Blockchain Technology in the Evolution of Humanity as a "Superorganism"

Martin Schmalzried[1]

1 School of Information and Communication Studies, University College Dublin, Dublin, Ireland; martin.schmalzried@ucdconnect.ie

* Corresponding author: Martin Schmalzried (martin.schmalzried@ucdconnect.ie)

**Abstract**

This article examines the broader societal implications of blockchain technology and crypto-assets, emphasizing their role in the evolution of humanity as a "superorganism" with decentralized, self-regulating systems. Drawing on a process philosophy approach grounded in Stiegler's "general organology" and further informed by related concepts such as Nate Hagens' "superorganism" idea and Francis Heylighen's "global brain" theory, the paper contextualizes blockchain technology within the ongoing evolution of governance systems and global systems such as the financial system. Blockchain's decentralized nature, in conjunction with advancements like artificial intelligence and decentralized autonomous organizations (DAOs), could transform traditional financial, economic, and governance structures by enabling the emergence of collective distributed decision-making and global coordination. In parallel, the article aligns blockchain's impact with developmental theories such as Spiral Dynamics. This framework is used to illustrate heuristically blockchain's potential to foster societal growth beyond hierarchical models, promoting a shift from centralized authority to collaborative and self-governed communities. The analysis, grounded in sense-making through a philosophical and biomimetical approach, and aims at providing a holistic narrative and view of blockchain as more than an economic tool, positioning it as a transductive technological seed for the evolution of society into a mature, interconnected global planetary organism.

**Keywords**: Blockchain; Bitcoin; Human development; Spiral dynamics; Decentralized governance; Societal impact

## 1. Introduction

The emergence of blockchain technology and crypto-assets, particularly Bitcoin, has sparked significant interest and debate regarding their potential impact on society, economics, finance and governance. Since the introduction of Bitcoin by the pseudonymous Satoshi Nakamoto [1], blockchain has been utilized for its ability to enable decentralized, transparent, and secure transactions without the need for intermediaries. This technological innovation carries the promise of revolutionizing various sectors, including finance, supply chain management, and even voting systems [2] all the while sparking many controversies including security risks, fraud and scams [3].

Despite the growing body of research on blockchain and crypto-assets, there is a notable gap in framing these technologies within a broader narrative that encompasses their societal implications and evolutionary significance. The innovative approach of this paper lies in contextualizing crypto-assets, blockchain, and Bitcoin within the larger narrative of humanity's development, through applied process philosophy, general organology and biomimetics. The epistemic angle is theoretical and speculative, aiming at exploring alternative narratives for the emergence of blockchain technologies which may broaden perspectives for future research.

The starting postulate around which this paper is built is the assumption that humans are an integral part of nature. While this may appear obvious, it has been a source of intense debate within philosophical thought, especially of so called Western Dualism, with René Descartes as an emblematic figure, separating reality into the res cogitans (thinking substance/mind: the human, thus thought to be somehow above/beyond nature) and res extensa (extended substance/matter) [4]. By reintegrating humans and thought/mind into nature, the endeavor of this paper is to treat human actions and activities through a biomimetical lens; assuming that human activity isn't disconnected from nature and thus presents similar patterns to those found in nature. One of the most prevalent patterns permeating all of nature is that of fractals: forms and patterns that repeat themselves across different scales while never reproducing themselves in exactly the same way. From the branching of trees, rivers, lungs, and neural networks to the recursive organization of ecosystems, fractals reveal how complexity can emerge through the repetition and variation of simple relational patterns. If humans are not outside nature, then human systems (economic, technological, political, and cultural, including blockchain) can also be understood as participating in this fractal logic. They reproduce, at another scale, dynamics and patterns already present in living systems at the noetic level (fractal patterns in and between ideas/concepts). For instance, the Zipf-Mandelbrot Law is a mathematical property of language which states that the frequency of any word is inversely proportional to its rank in the frequency table [5].

The aim of this paper is therefore not to reduce human activity to biology, but to explore how natural patterns can illuminate the structures and pathologies of human-made systems, particularly when they become detached from the broader ecological and relational conditions that sustain them, and especially through complexification of simpler patterns which create new emerging challenges. A leaf, a tree and a forest share common fractal properties and patterns. Yet a forest cannot be reduced to a leaf, in both its fundamental functioning and aspect. Nevertheless, this paper seeks to identify the core or fundamental patterns which may link micro somatic and social patterns to macro societal systems such as the financial system and blockchain.

Theories such as Simondon's philosophy of individuation, Bernard Stiegler's "general organology", Nate Hagens' concept of the human superorganism and Francis Heylighen's "global brain" hypothesis serve as a starting point for applying biomimetic and organological patterns to societal systems. Through such theories, new meanings and understanding of how technological developments such as blockchain can emerge, situating them in the broader context of the development of humanity as a whole, as a collective global organism. The general methodological approach takes its inspiration from Futures studies, which treats the future as a space for questioning dominant assumptions, expanding collective imaginaries, and exploring alternative socio-technical trajectories through original or uncommon paradigms, worldviews or assumptions.

Nate Hagens [6] describes humanity as a superorganism driven by collective behaviors that lead to exponential growth and resource consumption. This perspective highlights the challenges posed by the current trajectory of human development, which is characterized by a pursuit of economic growth often at the expense of environmental sustainability and social equity. Heylighen's "global brain" concept [7] envisions the internet and associated technologies as forming a collective intelligence that can process information and solve problems at a global scale. In both cases, the prevalent idea is that of humanity forming a global collective organism with its own sets of characteristics and features.

This paper explores how blockchain technology, in conjunction with other technological advances such as artificial intelligence (AI) and the metaverse, can play a key role in transitioning humanity from an unsustainable superorganism undergoing exponential and uncontrollable growth into a mature global entity with balanced growth and development (a balanced "global brain"). This paper posits that blockchain's decentralized nature aligns with the evolutionary shift toward more distributed and autonomous systems of governance and economic exchange. This shift mirrors the stages of human development outlined in Spiral Dynamics theory [8], which describes the evolution of human consciousness and societal structures through distinct levels, each with its own value systems and worldviews. Even though spiral dynamics has not been empirically validted to reflect humanity's evolution and development, this framework will be leveraged for its simple (if sometimes simplistic) developmental model which facilitates the identification of simpler fractal patterns in micro scale human development.

Spiral Dynamics provides a framework for understanding the parallels between individual human development stages, such as infancy, childhood, adolescence, and adulthood, and the evolution of human societies. For instance, the transition from egocentric (adolescent) stages to more collaborative and integrative (adult) stages reflects a shift from self-centered behaviors to a greater emphasis on community and systemic thinking. Applying this model, this paper examines how current societal structures are evolving and how technologies like blockchain fit within this broader evolution.

Furthermore, this paper delves into the symbolic significance of Bitcoin and other crypto-assets in the context of financial systems and monetary policy, exploring the idea that Bitcoin represents a form of "fail safe mechanism" or "savings", analogous to a young adult earning their first income and seeking autonomy from parental control (in this case, autonomy from centralized financial institutions and governmental bodies). This analogy extends to the challenges and opportunities inherent in this transition, including regulatory concerns, potential mismanagement, and the need for guidance and support to ensure a smooth shift toward decentralized systems.

Finally, this paper also addresses the critical question of how existing monetary policies and financial systems can adapt to the rise of crypto-assets by discussing the incompatibility between a debt-based monetary system reliant on continuous economic growth and a future where crypto-assets and stablecoins become mainstream.

In short, this paper aims to provide a comprehensive analysis that frames blockchain technology and crypto-assets within a broader narrative of human development and societal evolution. By integrating interdisciplinary theories and drawing symbolic parallels, this paper seeks to offer new insights into the potential pathways for a smooth transition between existing systems and novel and emerging systems. This holistic approach underlines the importance of viewing technological advancements not merely as tools for economic gain but as catalysts for deep societal transformation.

## 2. Theoretical Framework

### *2.1. Gilbert Simondon, Bernard Stiegler, process philosophy and general organology*

The ontological approach underlying this paper is grounded in process philosophy, which understands reality less as a collection of fixed and self-contained objects than as an ongoing process of becoming. Gilbert Simondon's philosophy of individuation is particularly relevant in this respect. Rather than taking individuals as already constituted entities, Simondon focuses on individuation as a process through which beings emerge in relation to a milieu (environment). Individuation is not the simple unfolding of a pre-given form, but the progressive resolution of tensions within a metastable field, through operations of transduction that generate new structures, norms, and relational possibilities [9]. At the societal level, technologies may therefore be understood not merely as external tools, but as mediating structures within the co-individuation of human beings, social institutions, and their environment. Bernard Stiegler further develops this intuition through the concept of general organology, according to which human life is shaped not only by physiological organs, but also by social organs such as law, language, education, and government, as well as technical organs such as writing, infrastructure, digital networks, and blockchain systems [10]. Societal systems are thus attached to nature through such an ontological grounding, as part of a larger individuating system which enfolds all of life including humans, technical objects and socio-technical systems.

This ontological foundation has important epistemological implications: the paper proposes an analogical and heuristic framework for identifying tensions within contemporary sociotechnical systems and exploring how these tensions may be transduced into new forms of coordination, governance, and monetary organization. In this sense, the analogies developed in the paper should not be understood as empirical proof, but as conceptual operators intended to expand the field of interpretation. Their relevance depends on whether they help clarify tensions that are already active within society and whether they open new possibilities for collective sense-making.

Victor Hugo illustrated this idea poetically through his famous quote: "Nothing is more powerful than an idea whose time has come". That is to say, certain ideas facilitate the transduction and reticulation of accumulated tensions or potential for transformation within a society than others. The aim of the paper is thus to present novel perspectives on the societal implications of the emergence of blockchain, based on a process philosophy ontology and coupled to Stiegler's "general organology", in the hopes of proposing a noetic seed that can situate blockchain, decentralization and self-governance through a sense-making exercise, which may facilitate a potential transition from centralization to decentralization.

### *2.2. Futures studies, questioning assumptions and expanding imagination*

Futures studies is an interdisciplinary academic field shaped by the work of scholars such as Fred Polak, Elise Boulding [11], Jim Dator [12], and Sohail Inayatullah [13]. The discipline has recently evolved from pure data driven predictive modelling to an expansion of future imaginaries via questioning one's assumptions about the future, one's worldviews, paradigms and unconscious beliefs. The present paper taps into this same approach to explore alternative futures scenarios for blockchain's integration in society, based on an original organological premise, and exploring the kinds of futures that can be derived from that assumption. Polak specifically linked a society's image of the future to its capacity for cultural renewal, arguing that societies flourish when they are animated by compelling and positive future images, and decline when such images lose their orienting and motivating force [11].

In this regard, the framework is informed by futures studies and process philosophy rather than predictive social science: it seeks to enlarge imagination and question common assumptions about the future belonging to more established academic paradigms for framing human-technology relationships such as techno-solutionism, techno-determinism, critical political economy, crypto-libertarianism or social constructivism. The following theoretical frameworks have been chosen specifically for their original organismic framing, building on the foundational process philosophy and general organology approach of the paper. The use of metaphors and analogies is further contextualized within such an approach, to consider future scenarios which would not be explored under more established paradigms.

### *2.3. Nate Hagens' concept of the human superorganism*

In his paper entitled "Economics for the Future - Beyond the Superorganism", Nate Hagens presents a narrative that frames humanity as a "superorganism," a collective entity driven by shared behaviors, values, and an insatiable appetite for growth and resource consumption. This concept draws parallels between biological organisms and human societies, suggesting that just as individual cells operate within a larger organism, individuals function within the broader context of society, contributing to collective outcomes that transcend individual intentions.

Hagens argues that at this stage, the human superorganism is characterized by unconscious patterns of exponential growth fueled by cultural narratives, technological advancements, and economic systems that prioritize immediate gains over long-term sustainability. This growth is often depicted as being out of control, leading to environmental degradation, resource depletion, and social inequities. The superorganism operates under the illusion of perpetual expansion, disregarding the finite nature of Earth's resources. The recognition of humanity as a superorganism emphasizes the need to understand the systemic drivers of our collective behavior. Hagens suggests that addressing global challenges requires a shift from individualistic perspectives to a holistic understanding of interconnected systems. This shift involves re-evaluating economic models, governance structures, and societal values to bring about a sustainable balance between human activities and the Earth's ecological limits. In this paper, blockchain and crypto-assets are presented as part of this process of re-evaluation and shift.

### *2.4. Parallel concepts in collective intelligence and societal evolution*

#### *2.4.1. The Francis Heylighen's "Global Brain" hypothesis*

Francis Heylighen introduced the concept of the "global brain," a metaphor for the collective intelligence emerging from the interconnectedness of individuals through technology, particularly the internet. The global brain represents a self-organizing network where information and knowledge are shared and processed collectively, enhancing the problem-solving capabilities of humanity as a whole. Heylighen posits that as communication technologies advance, they facilitate the development of a distributed cognitive system that mirrors the neural networks of a biological brain. This system allows for rapid dissemination of information, collaborative innovation, and adaptive responses to global challenges. The global brain is seen as an evolutionary step toward a more integrated and intelligent society. The global brain hypothesis aligns with the superorganism concept by emphasizing the collective aspects of human cognition and behavior. It suggests that technologies enabling decentralized communication and collaboration such as blockchain and crypto-assets can enhance the efficiency and adaptability of the global superorganism, and are a part of this global brain’s natural development.

#### *2.4.2. Ben Goertzel's views on artificial general intelligence and decentralization*

Ben Goertzel explores the intersection of artificial intelligence (AI), collective intelligence, and societal evolution [14]. He envisions the development of Artificial General Intelligence (AGI) that can understand, learn, and apply knowledge in a generalized way, much like a human being. Goertzel argues that AGI, integrated within decentralized networks, can significantly enhance humanity's cognitive capabilities.

Goertzel emphasizes the importance of decentralization in AI development, advocating for open, collaborative platforms that democratize access to AI technologies. He suggests that decentralized AI systems can prevent the concentration of power and promote a more equitable distribution of technological benefits. By integrating AGI into a global brain framework, humanity can arguably achieve higher levels of consciousness and problem-solving abilities. Furthermore, Goertzel highlights the potential of blockchain technology to support decentralized AI networks. Blockchain's secure, transparent, and distributed ledger systems can facilitate the coordination and incentivization of contributions to AI development. This synergy between AI and blockchain could accelerate the evolution of the global brain, leading to more sophisticated forms of collective intelligence.

All of the ideas above can also be linked to Teilhard de Chardin’s concept of the “noosphere”: the idea that, after the geosphere and biosphere, the Earth gives rise to a sphere of thought and collective consciousness, progressively intensified by language, technology, communication networks, and now potentially by AI, blockchain and other decentralized systems of coordination [15].

### *2.5. Spiral dynamics theory applied to societal and individual development*

To explore the parallels between individual human development and societal evolution, this paper leverages Spiral Dynamics theory, originally proposed by Clare R. Graves, and developed by Don Edward Beck and Christopher Cowan [8]. Spiral Dynamics is a psychological and sociological model that describes the evolution of human consciousness and cultural values through a series of stages, each characterized by distinct worldviews, motivations, and social structures.

Spiral Dynamics is used in this paper in a limited and methodological sense. It functions as a conceptual prism that helps organize an already established assumption of the paper: that certain patterns of development may recur analogically across different scales, from individual psychosocial development to broader societal and technological development.

The value of Spiral Dynamics for the present paper lies primarily in the simplicity of its staged vocabulary. By offering a sequence of broad developmental orientations, it facilitates the comparison between micro-level patterns of maturation, such as dependence, rebellion, autonomy, responsibility, and integration, and macro-level patterns of social organization, such as hierarchy, institutional authority, democratic contestation, decentralization, and distributed coordination. The framework therefore serves as a mediating schema between individual development and historical-social development. Its role is heuristic rather than evidentiary: it helps structure the analogical mapping, but it does not, by itself, demonstrate that societies necessarily evolve according to these stages.

This distinction is important because the application of developmental stages to societies is necessarily more hazardous than their application to individual development. Individual psychosocial development, although never perfectly linear, generally proceeds through relatively stable biological and social sequences. Civilizational and societal development, by contrast, is much more discontinuous, recursive, and fragile. Societies may develop complex institutions, collapse, regress, recombine earlier forms, or rebuild new configurations from the remnants of previous ones. Historical development is therefore not a smooth line but a series of attempts, failures, reticulations, and renewed individuations.

From a Simondonian perspective, these discontinuities can be interpreted as incomplete or unstable processes of collective individuation. A civilization may be understood as a transindividual configuration that temporarily resolves certain tensions between individuals, institutions, technologies, and their milieu. When such a configuration can no longer sustain those tensions, it may dissolve, collapse, or be transformed. However, the individuals who participated in that configuration do not simply return to a previous state. Their psychic and collective individuation has been altered by the social structures, institutions, techniques, and symbolic systems through which they lived. These transformations can then become the preindividual charge for later collective formations.

In this sense, individual and societal development should not be understood as separate processes, but as recursively co-individuating dynamics. Children do not develop in the same way in a small tribal group, a Roman city, a medieval village, an industrial nation-state, or a digitally networked society. Each social formation provides a different milieu for psychosocial development. In turn, the individuals formed within those milieus become capable of producing new social, technical, and institutional structures. Societal development shapes individual development, while individual development feeds back into societal development. This recursive loop is consistent with Simondon's understanding of individuation and transindividuation: the individual, the collective, the technical object, and the milieu co-emerge through relational tensions and their partial resolutions.

Spiral Dynamics simplifies this complex recursive process. It does not fully capture failed civilizational attempts and historical complexity, or the coexistence of multiple developmental logics within the same society. Nevertheless, its simplified structure is useful for the specific purpose of this paper, because it allows the organological and biomimetic argument to be presented as a comparative framework. It helps make visible the possibility that blockchain and decentralized governance may express a contemporary attempt to resolve tensions between centralization and autonomy, institutional authority and self-governance, inherited monetary systems and emergent forms of distributed coordination, as a recursive fractal projection of individual psychosocial tensions.

A further limitation of Spiral Dynamics is its Western-centric character. Its developmental sequence maps more easily onto the history of Western modernity than onto the plurality of social ontologies, cosmologies, and institutional forms found in Indigenous, African, Asian, or other non-Western contexts. However, within the scope of the present paper, it also explains why Spiral Dynamics remains analytically relevant. Blockchain, Bitcoin, decentralized finance, and DAOs emerged largely within technological, economic, and ideological contexts shaped by Western modernity: liberal individualism, market rationality, cyber-libertarianism, techno-solutionism, distrust of centralized authority, and the search for new forms of institutional coordination. The tensions surrounding blockchain are therefore deeply entangled with Western socio-technical imaginaries.

For this reason, Spiral Dynamics is not presented here as a universal anthropology of all cultures, but as a useful framework for interpreting a technology whose dominant narratives have been strongly shaped by the recursive co-individuation between Western socio-technical systems and Western psychosocial individual development.

The societal development stages, often represented by colors for ease of reference, include:

1. Beige (Survival/Safety): This foundational stage focuses on basic survival instincts and physiological needs. Societies at this level are concerned with immediate necessities, and individuals operate primarily on instincts.

2. Purple (Tribal/Animistic): Marked by a sense of community and mysticism, this stage values tradition, rituals, and the guidance of ancestral spirits. Social structures revolve around kinship and collective safety.

3. Red (Egocentric/Power Gods): Individuals assert themselves, seeking power, dominance, and immediate gratification. Societies may be ruled by strong leaders or warlords, emphasizing personal freedom without regard for others.

4. Blue (Absolutist/Mythic Order): Characterized by adherence to absolute truths, order, and stability. Societies develop structured institutions, laws, and moral codes, often underpinned by religious or ideological doctrines.

5. Orange (Achievist/Scientific Achievement): This stage values rationality, individualism, and progress. Societies encourage competition, innovation, and the pursuit of success through scientific and economic advancements.

6. Green (Communitarian/Egalitarian): Emphasizes community, relationships, and social responsibility. Societies prioritize equality, environmentalism, and consensus-driven decision-making.

7. Yellow (Integrative/Systemic): Individuals recognize the complexity of systems and seek to integrate knowledge from various disciplines. Societies focus on flexibility, sustainability, and holistic approaches to problem-solving.

8. Turquoise (Holistic/Globalist): Represents a global consciousness that transcends individual and collective interests. There is an emphasis on unity, interconnectedness, and the synergy of life systems.

Spiral Dynamics argues that individuals and societies can evolve through these stages, though not necessarily in a linear fashion. Each stage builds upon the previous ones, incorporating earlier values while expanding to encompass more complex perspectives. Each stage also creates its own institutions, which mirror its values and developmental characteristics.

**Table 1.** Comparative framework of individual and societal development stages

| Individual development stage | Perception of parents | Societal evolution stage | Perception of authority | Spiral Dynamics stage | Description |
|---|---|---|---|---|---|
| Infancy (0–2 years) | Parents perceived as mysterious, omnipotent beings; the world is magical and incomprehensible | Early Tribal Societies | Nature filled with spirits and gods; reliance on shamans and mystical figures. | Purple (Magical Animistic) | Focus on safety, tradition, and mysticism; strong family bonds; reality explained through myths and magic. |
| Early Childhood (3–6 years) | Parents seen as kings; admiration and acceptance of authority without question. | Formation of Monarchies | Kings and priests revered as divine or chosen by gods; authority is absolute. | Blue (Authoritarian Absolutist) | Emphasis on order, rules, and obedience; belief in one right way; institutions maintain stability. |
| Late Childhood to Adolescence (7–12 years) | Beginning to recognize parents' imperfections; questioning authority; desire for autonomy increases. | Rise of Democracies, census suffrage | Leaders elected but no longer revered; citizens critique and demand accountability. | Orange (Achievist Strategic) | Focus on individualism, achievement, and rationality; questioning traditional authority; pursuit of success and autonomy. |
| Adolescence (13–19 years) | Rebellion against parents; critical of authority; seeking independence while still reliant on support. | Demand for Rights and Social Movements, equal suffrage | Citizens advocate for civil rights and increased freedoms; challenge existing systems. | Green (Communitarian Egalitarian) | Emphasis on equality, community, and social responsibility; challenging existing systems for inclusivity. |
| Young Adulthood (20–29 years) | Establishing independence; navigating responsibilities; redefining relationship with parents. | Emergence of Decentralized Governance | Development of self-governing communities; use of blockchain and DAOs. | Yellow (Integrative Systemic) | Systems thinking emerges; recognition of complexity; seeking sustainable solutions. |
| Mature Adulthood (30+ years) | Fully independent; parents become advisors; individual contributes meaningfully to society. | Self-Governed Societies | Centralized governments make way for autonomous communities; collaborative problem-solving. | Turquoise (Holistic) | Global consciousness develops; holistic understanding and interconnectedness emphasized. |

# 3. Methods

### *3.1. Research Design and Futures-oriented Conceptual Approach*

This article adopts a conceptual, interdisciplinary, and futures-oriented research design. It does not aim to provide a systematic literature review, a quantitative empirical test, or a predictive model of blockchain adoption. Rather, it develops an abductive and interpretive synthesis aimed at broadening the socio-technical imaginaries through which blockchain technology can be understood.

The methodological orientation is informed by futures studies, particularly the idea that the future should not be approached only as an object of prediction, but also as a space for questioning assumptions, expanding imagination, and exploring alternative narratives. In this sense, the paper does not select its theoretical assumptions because they have been empirically demonstrated to be superior to competing paradigms. Instead, it selects them according to their conceptual fitness for operationalizing the paper’s ontological grounding in process philosophy, Simondonian individuation, and Stieglerian general organology.

The central methodological question is therefore whether an organological and process-philosophical framework can reveal tensions, analogies, and possible futures that are less visible within more established paradigms such as techno-solutionism, technological determinism, social constructivism, or critical political economy. The paper should therefore be read as a contribution to conceptual sense-making and futures-oriented imagination.

### *3.2. Data Collection and Materials*

The material used in this article was collected through a purposive and iterative literature-search strategy. Rather than attempting to exhaustively map all academic literature on blockchain, the search was organized around the conceptual assumptions developed in the paper. Sources were selected when they helped clarify, support, challenge, or contextualize the analogical correspondences explored in the framework.

Searches were conducted using academic databases and repositories such as Google Scholar, SSRN, or Scopus. The search relied on keyword clusters combining blockchain-related terms with the conceptual domains mobilized in the paper.

The search process followed a form of theoretical sampling. For each conceptual analogy, sources were gathered to assess whether similar patterns, tensions, or symbolic functions were already present in existing scholarship, policy debates, or public discourse. For example, when the paper explores the analogy between Bitcoin and bodily reserves, sources were selected that discuss Bitcoin as a store of value, a strategic reserve, a hedge against monetary instability, or an emergency asset in contexts of financial crisis. Similarly, when the paper examines blockchain as a possible organ of decentralized coordination, sources were selected from literature on DAOs, decentralized governance, regulatory sandboxes, and blockchain-based institutional experimentation, as well as on broad media perception and narratives around blockchain. Inclusion criteria were based on relevance to the paper's conceptual framework.

Because the paper is conceptual and exploratory, the material collected is not treated as a representative dataset. Instead, it serves as a set of conceptual anchors for developing and testing the plausibility of the proposed analogical framework. The aim is not to prove that the analogies are empirically exhaustive, but to examine whether they are theoretically coherent, interpretively useful, and capable of opening new research questions and avenues of thinking.

### *3.3. Analytical Procedure: Analogical and Transductive Synthesis*

The analysis proceeds through analogical and transductive synthesis. Analogical thinking is used here as a method for comparing processes of individuation across different scales. This approach is grounded in Simondon's philosophy of individuation, in which analogy can reveal structural operations shared by distinct domains without reducing one domain to the other. The paper therefore does not claim that society literally develops like an individual, or that blockchain literally functions as a biological organ. Rather, it asks whether similar relational tensions may be observed across biological, psychological, social, monetary, and technological processes.

The analytical procedure followed five main steps.

First, the paper defines its ontological premise: human societies and technologies are understood as ongoing processes of co-individuation rather than as fixed entities. From this perspective, blockchain is not treated merely as a technical instrument, but as a possible mediator in the evolving relationship between individuals, institutions, monetary systems, and collective forms of coordination.

Second, the paper identifies recurring tensions and developmental patterns across domains (biological, psycho-individual development, societal development). These include tensions between dependency and autonomy, centralization and decentralization, control and self-governance, growth and homeostasis, institutional authority and distributed coordination, and inherited systems and emerging socio-technical forms.

Third, the paper constructs analogical correspondences between individual development, social evolution, financial systems, and blockchain-enabled decentralization. These correspondences are organized in the comparative framework presented in Table 1, which functions as the central analytical matrix of the article. The table should therefore be understood as a heuristic device for exploring whether similar tensions can be observed across different levels of organization.

Fourth, the paper compares these analogical correspondences with selected literature and policy debates. For example, the analogy between Bitcoin and "fat" as a "bodily reserve mechanism" is compared with literature describing Bitcoin as a store of value, hedge, or strategic reserve. The analogy between parental authority and state regulation is compared with literature on regulatory sandboxes, collaborative regulation, and decentralized governance. This step is intended to prevent the analogies from remaining purely speculative by situating them in relation to existing debates.

Fifth, the paper distinguishes between three levels of claim: empirical evidence, conceptual analogy, and futures-oriented implication. Empirical evidence refers to existing technical, economic, or policy developments documented in the literature. Conceptual analogy refers to the interpretive correspondences developed between different domains. Futures-oriented implication refers to the possible scenarios or research questions that emerge from the framework. This distinction is important because the analogies are used to explore how blockchain may be interpreted as one possible expression of broader tensions within contemporary socio-technical individuation. Any predictive claims are extrapolated from taking the assumptions as true, and imagining what the implications would be on the future.

### *3.4. Methodological Limits*

The main limitation of this approach is that it does not provide empirical confirmation of the framework. The article does not claim that Spiral Dynamics, organology, or developmental analogies offer a universally valid model of human history. Nor does it claim that blockchain will necessarily produce more democratic, mature, or decentralized societies. The contribution is instead conceptual and exploratory: it proposes a framework through which blockchain can be interpreted as part of a broader process of socio-technical individuation, and exploring the implications of such a perspective.

The value of this framework should therefore be assessed according to its heuristic and futures-oriented usefulness: whether it clarifies tensions in current blockchain debates, whether it makes visible assumptions that are usually left implicit, whether it enhances the sense-making dimension in relating to technology and blockchain, whether it expands the range of imaginable futures, and whether it generates new hypotheses for future empirical research. In this sense, the paper aligns with futures studies by treating imagination not as a departure from scholarly rigour, but as a disciplined method for questioning dominant paradigms and exploring alternative socio-technical trajectories.

## 4. Comparative Framework exploration

To illustrate the fractal patterns in human and societal development, this paper presents a comparative framework that aligns stages of individual human growth with societal evolutions, perceptions of authority, and the corresponding Spiral Dynamics stages. This framework tends to show how individual psychological development (through the relation of an individual to his/her parents) mirrors societal transformations, particularly in the context of governance and the perception of authority (through the relation of a society to institutions).

See table 1 above.

### *4.1. Exploration of the framework*

1. Infancy and early tribal societies

Human level perspective: Infants perceive their parents as omnipotent and mysterious entities who control their environment in incomprehensible ways [16]. Their world is filled with wonder and magic, with no clear distinction between self and others.

Societal parallel: Early human societies interact with nature through animism and shamanism, believing in spirits and gods that influence their reality [17]. Shamans serve as intermediaries between the physical world and the spiritual realm.

Spiral Dynamics stage: Purple; Characterized by a magical-animistic worldview, strong family bonds, and reliance on traditions and rituals.

2. Early childhood and monarchies

Human level perspective: Children view their parents as authoritative figures or kings, accepting their decisions without question but recognizing some similarities between them and their parents [18]. They admire their parents and internalize rules.

Societal parallel: Societies evolve into monarchies with kings perceived as divinely appointed rulers. Authority is absolute, and obedience is expected [19].

Spiral Dynamics stage: Blue; Emphasizes order, conformity, and a strict hierarchical structure. Laws and institutions are established to maintain stability.

3. Late childhood to adolescence and democratic movements

Human level perspective: As children grow, they begin to notice their parents' imperfections and question authority. They seek greater autonomy and start forming their own opinions [20].

Societal parallel: The rise of representative democracy sees citizens no longer viewing leaders as infallible. There is increased scrutiny, demand for accountability, and a push for individual right [21].

Spiral Dynamics stage: Orange; Focuses on individualism, rationality, and the pursuit of personal success. Traditional authorities are questioned in favor of meritocracy.

4. Adolescence and social movements

Human level perspective: Teenagers often rebel against parental control, seeking independence while still relying on family support. They critique inconsistencies and advocate for personal freedoms [22].

Societal parallel: Social movements emerge, demanding civil rights, equality, and greater participation in governance. Citizens protest against perceived injustices and advocate for systemic change [23].

Spiral Dynamics stage: Green; Values community, equality, and environmental concerns. There is a focus on consensus-building and social responsibility.

5. Young adulthood and decentralization

Human level perspective: Young adults establish independence, manage responsibilities, and redefine relationships with their parents as peers or advisors [24].

Societal parallel: The emergence of decentralized governance models, such as blockchain technology and Decentralized Autonomous Organizations (DAOs), reflects a shift toward self-governance and reduced reliance on centralized authorities [25].

Spiral Dynamics stage: Yellow; Characterized by integrative thinking, flexibility, and an appreciation for systemic interconnectedness. Solutions are sought that are sustainable and adaptive.

6. Mature adulthood and self-governed societies

Human level perspective: Mature adults are fully independent and contribute meaningfully to society. They often guide younger generations and collaborate with others for common goals [26].

Societal parallel: Societies evolve into self-governed communities where whatever is left of centralized institutions such as governments facilitate rather than control. There is collaboration between authorities and citizens for problem-solving and a progressive shift to self-governing systems without any centralized intermediary.

Spiral Dynamics stage: Turquoise; Embodies holistic thinking, global consciousness, and an emphasis on the well-being of all life forms. Recognizes the interconnectedness of systems.

## 5. Humanity's development as a fractal pattern

The concept of fractal patterns suggests that structures and behaviors observed at one scale can be mirrored at other scales, exhibiting self-similarity across different levels of complexity [27]. In the context of human development and societal evolution, this implies that the stages an individual undergoes from conception to maturity may be thought of as an unconscious basic template or pattern for the broader developmental trajectory of human societies.

### *5.1. Symbolic parallels in finance and technology*

The evolution of financial systems evolves in parallel to the developmental stages of human growth and societal transformation. Symbolic parallels can be drawn between individual financial independence, the decentralization of economic systems, and the maturation of societal structures facilitated by technology such as blockchain. Other symbolic parallels can be drawn between the financial system and bodily systems within the human body, shedding an original perspective on the role of finance within humanity's superorganism. Such an approach is related to "general organology" whereby socio-technical systems and institutions are treated as organs within the human collective; representing exosomatic supports of psychic and collective individuation, involving thought, memory, labour, desire, attention, habit, and social coordination, as argued by Bernard Stiegler [10].

### *5.2. Financial independence and decentralization*

The transition from adolescence to adulthood is marked by achieving financial independence, which involves earning income, managing expenses, and making financial decisions autonomously [28]. This period is often challenging, as young adults may lack experience in financial management, leading to potential issues such as overspending or getting into debt. Beyond the strictly financial dimension, the move into shared accommodation also requires the development of new forms of interpersonal coordination. Young adults must learn to manage common resources as well as domestic responsibilities, set boundaries, and resolve conflicts without relying on the familiar hierarchical structure of the family household. In this context, relationships are no longer organized primarily around parental authority, but around interactions with peers who are theoretically situated on an equal footing. In practice, however, new tensions and power asymmetries emerge, as differences in income, personality, emotional maturity, among many other individual factors, can shape informal hierarchies within the shared living environment. However, within these interpersonal dynamics, parental intervention is rare (a parent stepping in to resolve disputes among cohabitants).

Similarly, the emergence of blockchain technology and cryptocurrencies represents a societal shift toward financial independence from traditional centralized institutions like banks and governments [1]. Early adopters of crypto-assets have encountered challenges including market volatility [29], regulatory uncertainty, and security vulnerabilities [30]. These obstacles reflect the problems associated with adopting new financial paradigms at a societal level, as a fractal projection at a higher level of complexity, of the problems associated to transitioning from adolescence to adulthood and gaining financial independence. Within the blockchain space, one can clearly see analogical challenges such as managing common resources, "domestic responsibilities" (who is in charge of updating the code, of public communication, of managing the treasury), set boundaries, and resolve conflicts, which often prompts reactions typical of parents coming to visit their young adults' apartment: "he/she is clearly not ready to live on his/her own", which takes the form of numerous regulatory initiatives meant to supervise the wild space of experimentation within the crypto-sphere. The European Supervisory Authorities and the ECB (European Central Bank) [31]–[33] similarly warn that many crypto-assets are highly risky and speculative, often promoted aggressively to consumers, and may fall outside existing protection frameworks. These obstacles can be read, within the analogical framework of this paper, as the societal-scale equivalent of the difficulties encountered during the transition from dependence to autonomy, and parental reactions to unresolved tensions and challenges encountered by young adults during their initial attempts at autonomy.

Yet these same experimentations as young adults are formative and lead to the emergence of responsible adults who can hold their own, who are no longer dependent on their parents, and who can "innovate" and improve upon the parental "template" pushing society to evolve forward, with new ideas and values which eventually, are considered commonplace. Democracy, for instance, as an ideal, was far from an acceptable "norm" a few centuries ago, and the beginnings of democracy implemented in the early 19th century weren't always successful. Yet nowadays, very few citizens would advocate returning to monarchies of divine right.

One clear example of this dynamic is the Ethereum DAO (Decentralized Autonomous Organization) hack. In its early imaginary, the DAO appeared as one of blockchain's most enthusiastic "young adult" promises: an organization that could coordinate collective action without "parental supervision" (courts, banks, public institutions). DAOs were presented as a new form of algorithmic governance, where rules, voting rights, treasury management, and execution would be embedded directly into smart contracts, replacing interpersonal trust and institutional mediation with transparent, neutral, programmable infrastructure [34], [35]. The DAO, launched on Ethereum in 2016, therefore became the poster child experiment of this imaginary: a decentralized organization in which investors could allocate funds collectively to projects through token-based voting, with transactions and settlements executed autonomously by code [36].

The irony is that this first major DAO experiment was almost immediately undone by the very technical autonomy it celebrated. In June 2016, a hacker exploited a smart-contract vulnerability and drained more than $50 million worth of ether from The DAO; the episode became a textbook example in Ethereum security scholarship, especially in analyses of vulnerabilities and smart-contract attack surfaces [37], [36]. The Ethereum community responded not by filing for bankruptcy, pursuing the attacker primarily through courts, or suing the software environment that made the exploit possible (as an existing regular company operating under the current centralized legal and institutional environment would do), but by resolving the crisis internally through a contentious hard fork that moved the affected funds into a new contract. This decision split Ethereum into two chains, Ethereum and Ethereum Classic, and exposed the tension between the rhetoric of decentralization and the practical concentration of protocol power: the outcome was not a formal democratic procedure in the conventional institutional sense, but a rapid emergency intervention coordinated by a relatively small set of technically and economically influential actors, then ratified through adoption by miners, exchanges, developers, and users. Later empirical work on Ethereum governance and DAOs supports this broader concern, showing that formally open governance can still be shaped by concentrated power, core development influence, token-weighted voting, and “inner power circles” [38], [39].

Read through the young-adult analogy, the DAO hack was less like a regulated corporation calling in courts, regulators, or bankruptcy administrators, and more like a crisis inside a shared apartment: the residents discovered that their supposedly self-managing household had produced a serious coordination failure, then improvised an in-house solution to restore order. The vulnerability was not simply “forgiven”; it became a formative lesson in smart-contract security and governance design, strengthening the ability to solve conflicts, tensions and security breaches through learning from mistakes. The same pattern appears later in Vitalik Buterin’s DAICO proposal, which sought to respond to ICO abuses by redesigning the token-sale mechanism so contributors could progressively release funds through a “tap” and potentially shut the project down by vote. This reflects a specifically young-adult mode of institutional learning: an attempt to build native mechanisms of accountability, self-limitation, and collective responsibility inside the new environment itself [40], [41].

### *5.3. Money as humanity’s circulatory system: Bernard Lietar’s perspective*

Bernard Lietaer, a renowned monetary theorist, compares the financial system to the circulatory system in the human body, with money acting as the blood that flows through economic "veins" to nourish various parts of society [42]. He argues that just as a healthy circulatory system is vital for physical well-being, a robust and diverse monetary system is essential for economic stability and resilience. Lietaer emphasizes the need for complementary currencies and decentralized financial structures to promote sustainability and adaptability in the face of economic crises, and likens financial crisis to cardiovascular diseases, whereby blood flow can be constricted which disrupts the “economy” (or in this sense, disrupts the coordination capabilities of individual human “cells” which, without money, can no longer participate effectively in the circulation of goods, services, information, and mutual obligations that sustains the broader social organism). In this sense, a financial crisis does not merely represent a technical failure of markets, but a blockage in society’s capacity for coordination, preventing its constituent “cells” (individuals, households and businesses) from exchanging energy, resources, and trust in ways necessary for collective vitality.

### *5.4. Money as a nervous system: Brett Scott's perspective*

Brett Scott offers another metaphor by likening money to the nervous system of a societal superorganism, where finance acts as the motor cortex that coordinates collective action [43]. In his article "Money is a Nervous System," Scott critiques traditional metaphors that compare money to blood flowing through an economic body. He argues that such metaphors are misleading because they suggest that money carries intrinsic value like nutrients in blood plasma, obscuring the true nature of finance.

Scott posits that money does not contain value itself but serves as an impulse that activates economic agents, much like nerve signals prompt muscles to move. He explains that value resides in people and their labor, not within the money. Monetary transactions are, therefore, not transfers of value but signals or information that mobilize resources and labor within the economy.

The financial system, acting as the nervous system, coordinates activities across the superorganism, enabling complex interactions and collaborations at scale. The central nervous system, represented by major financial centers and institutions, orchestrates large-scale economic actions through the issuance and management of money, where and how money should circulate in the economy, etc.

### *5.5. Combining both perspectives*

A more comprehensive understanding of the financial system may emerge by viewing it as a combination of both the circulatory and nervous systems. In this dual metaphor, money serves both as the signaling mechanism and the carrier of value. Just as the circulatory system transports oxygen and nutrients necessary for cells to perform their functions, money provides the means for individuals and businesses to engage in economic activities. Oxygen enables the execution of the nervous system's commands, sustaining cellular functions; similarly, money enables economic agents to access vital goods and services such as food, shelter, heat and so on.

Moreover, while the nervous system transmits information and coordinates actions, it requires the circulatory system to deliver the energy and nutrients that make such actions possible. In the context of the economy, the Internet could be likened to the neural network facilitating communication, while money acts as both the impulse (signal) and the medium (energy) that mobilizes resources.

This integrated perspective addresses the complexity of the financial system and its multidimensional role within humanity's superorganism. It acknowledges that money is not merely a carrier of value or a signaling mechanism but functions as both, enabling the "orders" of the economic system to be carried out while sustaining the activities of humans.

### *5.6. From centralization to decentralization*

While inside the womb, a baby's vital or biological resources depend solely on a centralized external organ: the placenta. This organ serves as the centralized intermediary between the mothers' resources and the baby (developing organism). While in gestation, the baby develops certain organs which will enable it to gain a certain degree of autonomy without maintaining the same degree of total dependency on his mother's resources and the placenta, which takes the form of breathing oxygen with his own lungs. The transition from one system to the other takes the form of cutting the umbilical cord, at which point, the baby takes his first breath. In order to ensure a rather smooth transition from one system to another, the baby stocks up on some fat, as the lungs and stomach are not yet fully mature and operational at birth.

While living in his/her parents' home, a child's access to societal resources depends solely on centralized external agents: the parents. Parents serve as the centralized intermediary between society's resources and the child (developing human individual). A child cannot walk into a store and help him or herself to goods/services without the intermediation of parents. While living in the parental home, the child develops certain skills (social skills, formal education etc) and undergoes certain developmental stages (toddler, child, teenager, adolescent, young adult) which enable the child to gain a certain degree of autonomy without maintaining the same degree of total dependency on his parents' resources and finances. In order to ensure a smooth transition from living at his parents' home to living on his own, a young adult will get a first student job and set some money aside. The transition from one relational dynamic to another takes the form of cutting the umbilical cord a second time, when the young adult moves out of the parental nest and gains full financial autonomy.

While living under the rule of governments, a society's access to the planetary resources depends solely on the affordances granted by centralized institutions which play the role of the symbolic "parent" to society. Governments serve as a centralized intermediary between the planet's resources and society (developing human collective). In order to access planetary resources, one must pay taxes, respect the law, hold property titles, ask for permits, and so forth. While living under the rule of governments, society matures in its ability to self-govern and act responsibly without needing centralized governments to enforce "civil" behaviours or collective decisions. Gradually, a society gains a certain degree of autonomy from governments which takes the form of adopting their own currencies, coordinating their actions via DAOs (decentralized autonomous organisations) or forming communities which reclaim sovereignty over activities formerly performed by the state. In order to ensure a smooth transition from living under the rule of centralized governments to self-governed communities, and stop relying on centralized financial institutions (fiat money, or the debt-based monetary system), a society will start to set some money aside (Bitcoin) to weather the transition from one financial system to another.

In this instance, there are a number of parallels between Bitcoin and "fat" that the baby stocks up on before birth (as in the first fractal level above). In order for a baby (global organism) to store fat, it needs to spend extra "energy" in order to convert food into a molecule directly assimilable by the body (the individual cells). In order for society to store "value", it needs to spend extra energy in order to convert electricity into an energy which is directly assimilable by humanity (individual humans)1. In case of a global financial collapse, arguably, the only "system" which is relatively neutral (not under the control of a centralized authority or government) and globally accessible, at this time, is Bitcoin. Any individual with an internet connection can create a Bitcoin wallet and start transacting (receiving/sending Bitcoin), which is exemplified in cases where currencies and the financial system have failed, such as Lebanon [44]. During times of economic uncertainty or systemic transitions between financial systems, Bitcoin can serve as a stabilizer. Just as fat reserves provide energy during periods of scarcity, Bitcoin offers a decentralized asset that can preserve wealth and ensure that major necessary transactions can still take place when traditional financial systems are under stress [45]. As with any organism, exchanges must continue. The same holds true for humanity at the global scale. If exchanges were to suddenly freeze due to a collapse of the financial system, humanity would no longer behave like a collective global organism, but would revert back to fragmented, isolated communities operating independently, much like cells that lose their cohesive functioning when an organism's circulatory system fails.

Furthermore, as humanity progresses towards a more interconnected and technologically advanced society, the role of artificial intelligence (AI) becomes significant. AI could be envisioned as the "cognitive facilitator" of the global superorganism, processing vast amounts of data (the sensory input generated by humans) and generating coherent strategies (the motor output) that guide collective actions [14]. This analogy emphasizes that while money (as both nervous and circulatory systems) facilitates the functioning of the economic body, AI represents the cognitive processes that can potentially bring higher levels of decentralized coordination and create a meta-coherence that cannot be achieved with traditional centralized top-down institutions. The biomimetic equivalent is that of homeostasis in a biological organism, or automated processes which maintain the bodily environment in an ideal or optimal biological/biochemical state: mechanisms for adjusting bodily temperature, pH levels, blood pressure, sugar levels, etc. The latest paper published by Yann Lecun advocates for SAI (Superhuman Adaptable Intelligence) which could provide the cognitive architecture for planetary homeostasis: specialized intelligences which monitor and regulate distinct ecological, technical and social functions, while shared world models coordinate their interactions across the whole system. [46]

[1] Assimilable in the sense that humans can transact with it, and thus set human individual "cells" into motion, enabling global decentralized coordination. Energy in the form of electricity cannot easily be used for payment/transactions. Bitcoin, however, can. The process of Bitcoin mining can in this case be assimilated to a metabolization of "raw" energy into energy that can directly set humans into motion.

These analogies are not meant to be taken at face value, but serve the purpose of offering an alternative narrative around the ethical and practical implications of evolving financial systems to accommodate blockchain,crypto-assets and Artificial Intelligence as a specialized stabilizing homeostatic layer. For instance, the concept of a universal basic income (UBI) aligns with the circulatory system metaphor, where each "cell" (individual) receives the necessary "oxygen" (financial resources) to survive and function without stringent controls or conditions [47]. While transitioning to such a system, Bitcoin and other crypto-assets can serve as a “temporary bridge” much like the example of “fat” discussed above.

The recent move by various governments around the world to set up a “Bitcoin strategic reserve” is directly tied to the symbolic role of Bitcoin as a "store of value" during transitional periods [48], symbolically reflecting the way organisms store fat to prepare for significant changes. By accumulating Bitcoin reserves, these governments aim to safeguard their economies against potential instabilities in the traditional financial system, much like how a baby relies on fat reserves during the shift from placental nourishment to independent feeding, or how a young adult sets money aside before moving out of his/her parents’ home.

This strategic move illustrates the recognition of Bitcoin's potential to act as a financial stabilizer in the face of economic uncertainty. It aligns with the broader transition from centralized to decentralized systems, where reliance on traditional fiat currencies and centralized financial institutions is gradually diminishing.

### *5.7. Parental reactions and regulatory responses*

In the context of individual development, parents may react to their child's pursuit of financial independence with concern or attempts to maintain control, fearing that the child is unprepared for the responsibilities [49].

Analogously, governments and regulatory bodies often respond to the rise of decentralized finance with restrictive measures, citing risks such as fraud, money laundering, and threats to financial stability [50]. These apprehensions are not unfounded. The decentralized and permissionless nature of some crypto-assets presents challenges, including susceptibility to fraud, market manipulation, and security breaches, latest of which, the scandals provoked by meme coins [51]. Early adopters may suffer losses due to inexperience or insufficient safeguards, much like young adults who might make imprudent financial choices without proper guidance, or in the same way an unsupervised baby puts anything into his mouth. However, adopting a punitive approach may be self-defeating, in the same way as a parent seeking to forcefully regain control over his/her young adult, which may inadvertently precipitate their desire for independence and autonomy, albeit experienced in a chaotic way.

### *5.8. Navigating the Transition*

A supportive approach in both parenting and societal governance can facilitate smoother transitions to independence. Parents who provide guidance without exerting overbearing control help young adults develop competence, confidence, and responsibility [52]. They recognize the importance of allowing their children to make mistakes and learn from them, fostering growth and maturity. This is valid at the micro level (a baby that is allowed, within reason, to chew on a variety of objects develops a more resilient immune system) as well as the mezzo level (a young adult that is allowed, within reason, to manage his own money and make financial mistakes develops more overall resilience).

Similarly, governments can adopt a collaborative stance toward the evolving financial landscape by:

- • Creating regulatory sandboxes: Allowing innovators to test new financial products and services in a

controlled environment, enabling regulators to understand emerging technologies and develop appropriate frameworks [53].

- • Promoting best practices: Encouraging the development of industry standards and self-regulatory organizations that can establish guidelines for ethical behavior and risk management within the DeFi space [54].
- • Engaging with stakeholders: Involving technologists, entrepreneurs, consumers, and other stakeholders in the policymaking process to ensure that regulations are informed, balanced, and supportive of the transition from centralized to decentralized systems [55].
- • Providing free auditing services for new blockchain initiatives: Establishing government-funded programs and validate artificial intelligence use-cases that offer security audits, code reviews, and smart contract verification services to early-stage blockchain projects, reducing barriers to entry while ensuring baseline security standards and protecting users from preventable vulnerabilities. These services can help prevent technical failures and security breaches that could harm users and damage confidence in the broader DeFi ecosystem, while also creating a knowledge base of common vulnerabilities and best practices that can benefit the entire industry [56].
- • Revising existing systems and regulations: Initiating comprehensive reforms of fundamental financial and regulatory frameworks to accommodate the paradigm shift introduced by permissionless blockchain technology. This includes reimagining core mechanisms like money creation (moving from debt-based money creation to alternative monetary systems such as the Relative Theory of Money), taxation frameworks, and regulatory approaches. Rather than forcing blockchain innovation to conform to legacy systems, this approach recognizes the need to fundamentally redesign these systems to leverage the unique properties of blockchain technology while ensuring public benefit and stability. By adopting such measures, governments can help bridge the gap between traditional financial systems and emerging decentralized models. This approach aligns with the maturation process described in Spiral Dynamics, where societies transition from authoritative structures (Blue) to more participatory and integrative systems (Yellow and Turquoise).

The shift toward decentralized governance can occur through various pathways, each presenting unique challenges. An abrupt transition involves a rapid move away from centralized authority, which can lead to instability and resistance from existing institutions unprepared for sudden change. Alternatively, a gradual independence allows for a smoother transition, where governments and citizens collaborate to delegate responsibilities progressively. This approach can mitigate potential disruptions and foster mutual adaptation to new governance models. The same pattern applies to a young adult seeking independence from his/her parents: a smooth transition is preferable to an abrupt transition, whereby the young adult has to manage on his own with no parental support.

To ensure a stable and beneficial transition, governments can adopt strategies that support citizen-led initiatives. One such strategy is the delegation of competencies, gradually transferring responsibilities like infrastructure maintenance, environmental management, and social services to DAOs or community organizations, all the while maintaining a presence to monitor the outcomes in order to step in should these experiments turn sour [57].

## 6. Implications for monetary policy

The rise of crypto-assets and decentralized finance (DeFi) presents significant challenges to traditional monetary policy and existing financial systems. The coexistence of fiat currency and widespread crypto-assets could lead to an artificial multiplication of the monetary supply, disrupt central banking monetary transmission mechanisms, and impact the lending capacity of the legacy banking system, potentially causing economic instability [58]. This concern arises because crypto-assets like Bitcoin may serve as alternative means of payment and stores of value, operating parallel to traditional currencies without following the same logic in terms of issuance/creation and circulation.

In a debt-based monetary system, money is primarily created through the issuance of loans by banks, which requires continuous economic growth to service the interest on that debt [59]. The introduction of crypto-assets disrupts this model by providing alternative channels for transactions and value storage that are independent of central banks and traditional financial institutions. If crypto-assets become widely accepted as means of payment, the effective monetary mass increases without corresponding controls, potentially leading to inflationary pressures and undermining the effectiveness of monetary policy. In essence, money can now circulate twice. For instance, if someone sells 1000 tokens of a given crypto-asset which has been mined or airdropped (created out of thin air) for 1000€, and those same tokens are accepted widely as payment for goods and services, then the person who sold the tokens is now in possession of 1000€ to spend, while the person in possession of the 1000 tokens can also spend those into the economy. This isn't the case with traditional goods. When buying 1000€ worth of furniture, one cannot utilize the furniture as means of payment. In this sense, Bitcoin cannot be viewed as "digital gold", as one cannot walk into a bakery and scratch of a few miligrams of a gold bar to pay for bread. In order to use gold as a "means of payment", one must first sell it against an accepted (fiat) currency, or at least issue special "bonds" that are redeemable for gold[2]. Bitcoin, on the other hand, could theoretically be used as a means of payment as its value can be broken down into subunits, and also thanks to innovations such as the lightning network allowing for near-free and fast transactions [60]. However, using Bitcoin as means of payment can only be a temporary fix, much like relying on bodily fat, for obvious reasons: Bitcoin's deflationary characteristics. For payments, only a currency which manages to maintain its purchasing power over time can qualify as money or as a medium of exchange, otherwise the incentive to either hold it (in case of deflation) or get rid of it (in case of hyperinflation) distorts economic activities and consumer behaviour.

Moreover, the volatility inherent in many crypto-assets introduces additional risks. Fluctuations in the perceived value of crypto-assets can distort the balance between the monetary mass and the goods and services available in the economy. This instability could complicate the central bank's ability to manage inflation and maintain economic stability. While these effects have not been seen yet given the marginal adoption and weight of crypto-assets within the larger financial system, this could become a problem if the rate of adoption continues and crypto-assets become more and more accepted and recognized as means of payments.

Furthermore, the rise of stablecoins (crypto-assets pegged to traditional currencies or assets) adds another layer of complexity to the monetary system [61]. Stablecoins can circulate alongside fiat currency, effectively doubling the money in circulation if not properly accounted for (for instance, in case a government issues fresh bonds which are bought by a stablecoin issuer, thereby multiplying the monetary mass). The way stablecoin issuers manage underlying assets, such as investing customer deposits in bonds, could lead to the same money circulating twice in the financial system, exacerbating the tensions identified above[3].

The incompatibility of a debt-based monetary system with the growing use of crypto-assets points to the necessity to undergo a similar transition to the cutting of a baby's umbilical cord at birth, transitioning from a centralized "system" which enables growth of the organism (the debt-based monetary system, which is perfectly designed to incentivize growth) to another system where the logic is no longer based on growth but on adequacy between the organisms' needs, based on the organisms' actions. While a baby inside the womb grows steadily, with energy consumption going up, a baby outside of the womb is equipped with mechanisms allowing to modulate energy consumption based on the need (whether the baby is sleeping, crawling, lying down etc), via its lungs (taking in more or less oxygen) or stomach (eating more or less food). In effect, this means transitioning from a logic of exponential growth, to a logic of maturation and adequate fluctuation of resource allocation based on "needs".

[2] This would not apply to tokenized gold such as PAXG (Ethereum blockchain), which can then be fractionalized and effectively be used as payment.

[3] In this scenario, a stablecoin mints 1000 units of euro pegged tokens, for instance and sends them to the wallet of a European consumer. The consumer can now spend those 1000 tokens in place of fiat money. The stablecoin issuer then buys a freshly issued short term bond from the government which generates yield against the 1000€ received from the consumer. The government now has 1000€ to spend into the economy.

To this effect, a number of prominent economists have been advocating for a shift towards some form of "full reserve" monetary system such as Michael Kumhof which revisits the "Chicago Plan" (post-depression proposal to revive the economy) [62]. Full reserve banking has also seen more interest from civil society organizations such as Positive Money, advocating for more democratic control over money [63]. In this sense, exploring alternative monetary theories such as the Relative Theory of Money by Stéphane Laborde may be warranted, as these novel monetary theories rely on other mechanisms besides the debt-based creation/destruction cycles to stabilize a currency's value over time [64]. It is a form of full reserve banking system, but where monetary creation is equally distributed to each citizen. Our current debt-based monetary system is ill-equipped to weather erratic fluctuations in economic activity, preferring steady production/consumption patterns coupled to steady and predictable growth. The challenge is thus to think of monetary systems which mimic the same properties as the adaptable organic fluctuation of a baby's energy input and consumption after birth.

There are many more patterns one could explore beyond blockchain and the financial system which could be explored through the same organological and biomimetic framework, such as the energy transition, sustainable development, pollution and circular economics. However, these topics are outside the scope of the following paper.

## 7. The S-curve and post-growth economics

The relevance of the S-curve is that it offers a more precise biological image of maturity than either indefinite expansion or static equilibrium. Many natural growth processes begin with a slow phase, accelerate rapidly, and then gradually decelerate as the organism reaches a more mature form [65]. This can be observed in the growth of plants, in fetal development, and in the broader trajectory of human development before and after birth. Growth does not stop after birth, but its character changes: it becomes less about the rapid multiplication of mass and more about regulation, differentiation, coordination, learning and adaptation.

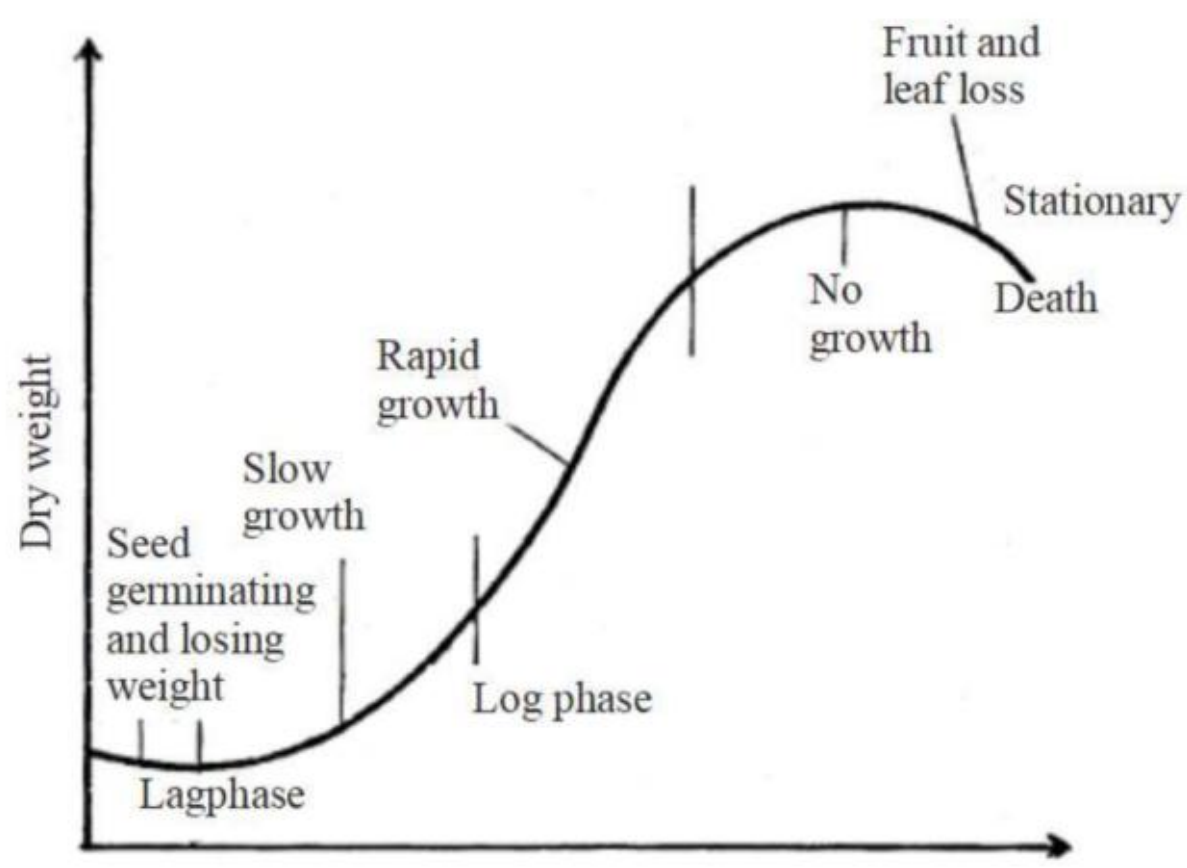

**Figure 1.** Plant growth as a sigmoid curve. Source: Studocu, "Growth curve" [66].

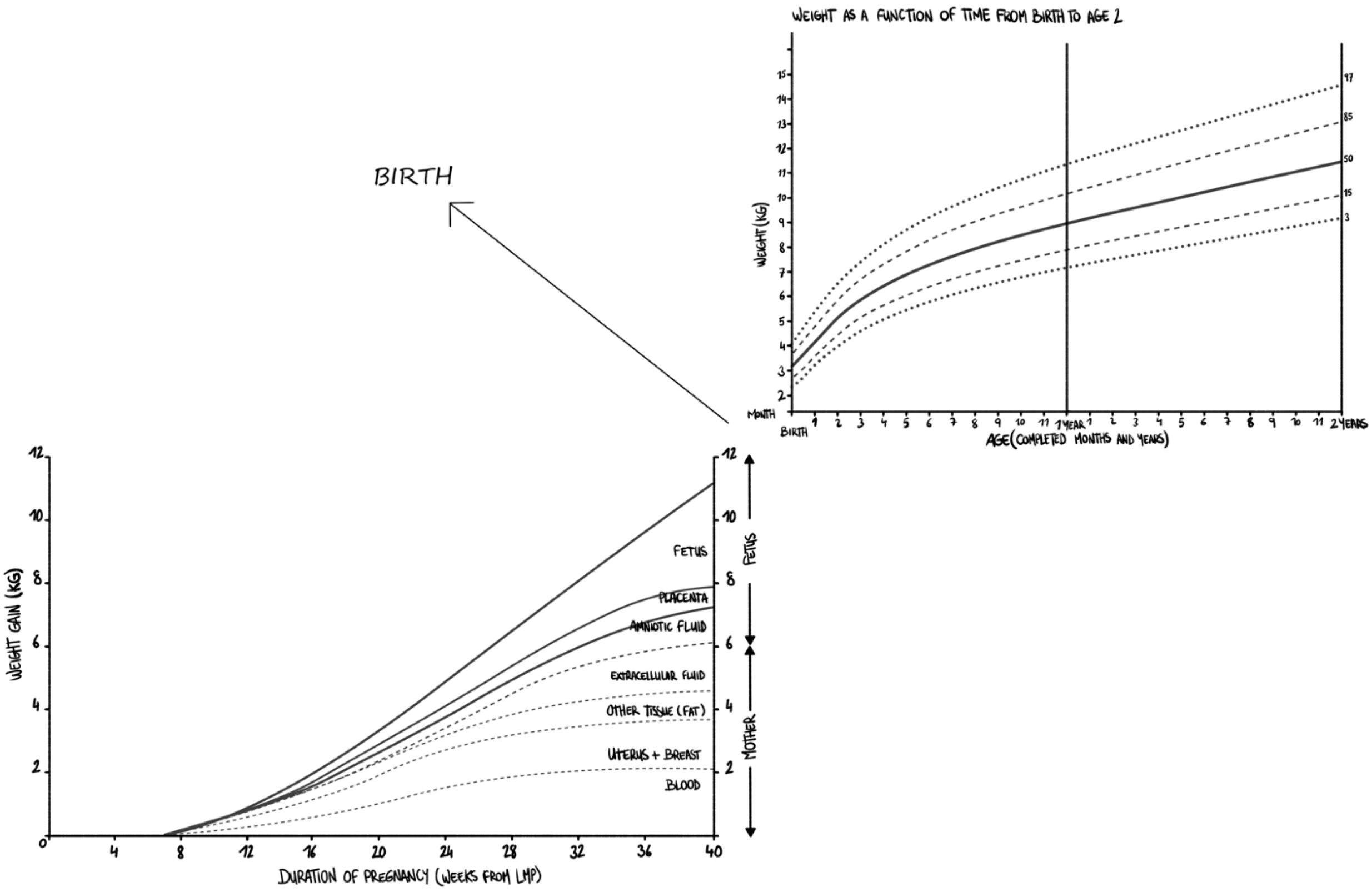

**Figure 2.** Fetal and early postnatal growth curves.

Source: Schmalzried, Mother Earth's final push: The story of humanity's gestation [67].

The S-curve growth pattern, incidentally, is close to what many so-called "post-growth" economists are calling for [68]. Read through the framework developed in the previous sections, this should not necessarily be understood as an artificial imposition of post-growth on an otherwise infinitely expanding economy. It may instead be interpreted as a natural transition from an early phase of expansion toward a phase of maturation. Human societies, like organisms, cannot remain indefinitely in an embryonic mode of development.

However, the analogy also shows why this transition must be handled carefully. A plant does not mature by being prevented from growing; it matures by reorganizing its growth into more complex forms. Similarly, a baby does not survive birth by remaining attached to the placenta in a state of controlled dependency. Cutting the umbilical cord marks a radical shift in the mode of resource distribution: the organism must move from centralized nourishment through the placenta to breathing, digestion, circulation and homeostatic regulation. In societal terms, the question is therefore not simply whether humanity should "grow" or "stop growing," but whether it can shift from a debt-based system structurally oriented toward expansion to a monetary and technological system oriented toward dynamic adequacy.

A word of caution, however, in contemplating these fractal parallels. While these transitions appear to be fractally informed, they may fall short of the magnitude of the transition symbolized by "birth". When reading recommendations by some "post-growth" or "de-growth" economists, one wonders if this represents a birthing of humanity, or rather, keeping humanity in a artificial controlled stasis "inside" mother Earth's planetary womb, compressing the umbilical cord (reducing oil consumption), and only swallowing as much amniotic fluid as the mother/placenta can renew (circular economy). This may seem appealing; however, any biologist will caution against such an attempt. In nature, stasis does not exist. Nor does infinite exponential expansion.

As argued above, the Relative Theory of Money is particularly relevant here because it shifts money creation away from debt-driven expansion and toward a regular, symmetrical flow distributed across the monetary community. It therefore corresponds less to a logic of accumulation than to a logic of circulation, allowing the financial system to modulate flows according to real needs and ecological limits.

Failure to follow the S-curve could result in uncontrollable cancerous growth of the human collective. Such an expansion would represent a form of cancerous development, rather than reaching a state of developmental maturity. Cancer has been described as an interruption of maturation, where cells proliferate instead of differentiating. This means, in effect, replicating without evolving, or in this case, capitalism as a system replicating a growth model without "differentiating" into a mature economic organ of collective regulation, capable of modulating economic flows according to real needs and ecological limits [69].

## 8. Discussion

When viewed through the lens of fractal patterns, general organology and biomimetics, the emergence of blockchain and decentralized technologies takes on a quasi-teleological dimension, not necessarily imposed or piloted from "above", but immanent to the specific tensions emergent from the relational dynamics prevalent in humanity, today. Heuristically, one could view blockchain as a Tetris block (informational/noetic singularity or seed) which is uniquely shaped to reticulate the accumulated tensions of contemporary society: distrust in centralized institutions, dependence on hierarchical intermediaries, the fragility of global financial systems, the desire for autonomy, and the need for new forms of coordination at planetary scale, mirroring a similar "need" in individual psychosocial development symbolized by the transition from adolescence to adulthood. Its significance would therefore not lie merely in its technical novelty, but in its capacity to crystallize, absorb, and reorganize pre-existing tensions into new socio-technical forms, informed by accumulated successful transductions of micro-societal tensions.

Rather than treating socio-technical systems as metastable transindividual structures disconnected from processes of individuation at the human level, this paper argues that collective individuation draws upon historically accumulated patterns of micro-social transduction. Successful resolutions of tension at smaller scales (within the body, the psyche, the family, the household, or the community) can become analogical templates for larger processes of social and technological organization. Individuation, in this sense, permeates multiple scales: it is not confined to the individual, but reverberates (or resonates, to use a Simondonian term) through the relational structures that connect individuals to groups, institutions, technical systems, and broader milieus. The possibility of more harmonious collective dynamics therefore depends, at least heuristically, on the capacity to transpose patterns of coordination, differentiation, and integration from one scale to another. Just as bodily life depends on complex relations between differentiated organs, and psychosocial life depends on the complex relations between differentiated individuals, societal life may be understood as a higher-order attempt to organize relations between partially autonomous yet mutually interdependent transindividual collectives, which takes the form of specific metastable social and technical/artificial organs. From this perspective, blockchain can be interpreted as one such transductive attempt: a technical form through which tensions around authority, trust, autonomy, and coordination are reorganized at the societal scale. Blockchain, in this regard, is but one possible technical manifestation of a broader organological phenomenon that could express itself through other forms, such as revisiting cooperative structures and other self-organizational systems.

Based on the papers' framework, a potential transition from centralized systems to decentralized system will take some time. While a baby's transition can take minutes (abruptly cutting the umbilical cord) to a few days (in case one leaves the placenta attached as in Lotus birth) [70], and while a young adult's transition can take a few days to a few months or even years, humanity's transition from centralized systems to decentralized ones could take a several decades, and follow a similar dynamic as previous major transitions such as transitioning from monarchies of divine right to representative democracy: highly diverse imperfect and iterative, reflecting each human society's cultural and institutional singularity, as transpires from the recent diverging regulatory attitude towards crypto-assets and stablecoins between the US and the EU [71].

Such a conclusion stems from the premise that humanity isn't disconnected from nature, but replicates at its own scale patterns found in nature, following a teleonomic trajectory. I propose the term morphonoetics to capture this research angle, derived from the term "morphogenesis" in biology [72] which studies the biological processes that cause an organism to develop its shape, and "noetic," relating to concepts/ideas, thereby defining the study of how the structural forms of human civilization evolve in resonance with human cognitive or noetic development.

In this framework, morphonoetics suggests that the "shape" of our social and financial institutions is not arbitrary, but a physical expression of metastable resolutions of societal tensions unique to our current stage of cognitive and spiritual development. Just as a developing embryo undergoes physical folding and differentiation to accommodate its increasing complexity, humanity is currently "folding" into decentralized networks to accommodate a level of global interconnectedness and information density through novel relational tensions which centralized, hierarchical structures may no longer contain or manage.

In contemplating a potential transition from centralized systems to decentralized ones, the key question is how to approach it: via an abrupt transition or a smooth and gradual one? There are many alternatives offered to us, mirroring the richness and diversity in ways young adults seek autonomy from their parents. It could lead to struggles as various stakeholders, actors and institutions resist change by clinging to centralized structures in resonance with the fractal pattern of a parent struggling to control a young adult in search of autonomy and independence. Or such a transition could be envisaged to be progressive and managed, through a gradual delegation of powers from centralized institutions to decentralized or self-governed systems (for instance, delegating road repair to decentralized citizen-led systems while maintaining centralized systems as a fallback mechanism). The aim of this paper is to shed light on these potential future scenarios, bring them into conscious awareness, in order to choose how to navigate such a transition consciously (should the pattern be verified).

Embracing decentralized systems with awareness enables a smoother transition, applying the developmental trajectories identified in Spiral Dynamics and aligning with frameworks like the Relative Theory of Money, which advocates for an adaptive, resilient financial system that distributes value sustainably and equitably. In this approach, governments and institutions act as facilitators rather than controllers, supporting the shift through collaborative regulation, innovation, and open engagement. By choosing to navigate this evolution with foresight, we have the opportunity to edge towards a decentralized future that respects both our collective interdependence and individual autonomy, enabling humanity to grow into a mature, cohesive superorganism aligned with the needs and values of a global society, and steering clear of the self-destructive path that the superorganism is currently on, as argued by Nate Hagens. A healthy parent takes pride in raising his/her child to the point where they become fully autonomous and self-governing adults. Governments and current centralized institutions of power should take pride in raising a society to the point where it is capable of self-governance.

While this framework does not present ready-made solutions to humanity's current challenges, it offers broader sense-making by shifting our perspective from a crisis of collapse to a crisis of transition, thereby providing for a much needed metaphysical answer to today's meaning crisis [73].

## 9. Conclusion

This article argues that blockchain technology and crypto-assets can be interpreted as part of a broader societal transition from centralized dependency toward decentralized self-governance. By reading blockchain through the lenses of transindividuation, general organology, the "human superorganism", the global brain, Spiral Dynamics, and monetary-system metaphors and analogies, the article frames decentralization not simply as a technical innovation but as a potential developmental shift in collective organization. Future research could test this conceptual framework's validity by examining emerging societal patterns from decentralized autonomous organizations, stablecoin ecosystems, regulatory sandboxes, and other blockchain-enabled governance arrangements which further resonate with the theoretical framework proposed above.

**Funding**
This research received no external funding.

**Data and Code Availability Statement**
No new datasets or code were generated or analysed during the current study. The sources supporting the conceptual analysis are listed in the References section.

**Conflicts of Interest**
The author declares no conflict of interest.

**Patents**
The author declares that there are no patents related to this study's methods or tools.

## Abbreviations

AGI – Artificial General Intelligence
AI – Artificial Intelligence
DAO – Decentralized Autonomous Organization
DAOs – Decentralized Autonomous Organizations
DeFi – Decentralized Finance
SSRN – Social Science Research Network
UBI – Universal Basic Income